\documentclass[aip,pop,preprint,superscriptaddress]{revtex4-1}

\usepackage{graphicx}
\usepackage{subfig}
\usepackage{upgreek}
\usepackage{caption}

\begin{document}

% Use the \preprint command to place your local institutional report
% number in the upper righthand corner of the title page in preprint mode.
% Multiple \preprint commands are allowed.
% Use the 'preprintnumbers' class option to override journal defaults
% to display numbers if necessary
%\preprint{}

%Title of paper
\title{Simulation study of a passive plasma beam dump using varying plasma density}

% repeat the \author .. \affiliation  etc. as needed
% \email, \thanks, \homepage, \altaffiliation all apply to the current
% author. Explanatory text should go in the []'s, actual e-mail
% address or url should go in the {}'s for \email and \homepage.
% Please use the appropriate macro foreach each type of information

% \affiliation command applies to all authors since the last
% \affiliation command. The \affiliation command should follow the
% other information
% \affiliation can be followed by \email, \homepage, \thanks as well.
\author{Kieran Hanahoe}
\email[]{kieran.hanahoe@postgrad.manchester.ac.uk}
\author{Guoxing Xia}
\author{Mohammad Islam}
\author{Yangmei Li}
%\homepage[]{Your web page}
%\thanks{}
%\altaffiliation{}
\affiliation{School of Physics and Astronomy, University of Manchester, Oxford Road, Manchester M13 9PL, United Kingdom\\}
\affiliation{Cockcroft Institute, Sci-Tech Daresbury, Keckwick Lane, Daresbury, Cheshire WA4 4AD, United Kingdom}

%Collaboration name if desired (requires use of superscriptaddress
%option in \documentclass). \noaffiliation is required (may also be
%used with the \author command).
%\collaboration can be followed by \email, \homepage, \thanks as well.
%\collaboration{}
%\noaffiliation
\author{\"Oznur Mete-Apsimon}
\affiliation{Department of Engineering, Lancaster University, Bailrigg, Lancaster LA1 4YW, United Kingdom\\}
\affiliation{Cockcroft Institute, Sci-Tech Daresbury, Keckwick Lane, Daresbury, Cheshire WA4 4AD, United Kingdom}

\author{Bernhard Hidding}
\affiliation{Department of Physics, University of Strathclyde, Richmond Street, Glasgow G1 1XQ, United Kingdom}
\affiliation{Cockcroft Institute, Sci-Tech Daresbury, Keckwick Lane, Daresbury, Cheshire WA4 4AD, United Kingdom}

\author{Jonathan Smith}
\affiliation{Tech-X UK Ltd., Sci-Tech Daresbury, Keckwick Lane, Daresbury, Cheshire WA4 4FS, United Kingdom}

\date{\today}

\begin{abstract}
A plasma beam dump uses the collective oscillations of plasma electrons to absorb the kinetic energy of a particle beam. In this paper, a modified passive plasma beam dump scheme is proposed using either a gradient or stepped plasma profile to maintain a higher decelerating gradient compared to a uniform plasma. The improvement is a result of the plasma wavelength change preventing the re-acceleration of low energy particles. Particle-in-cell simulation results show that both stepped and gradient plasma profiles can achieve improved energy loss compared to a uniform plasma for an electron bunch of parameters routinely achieved in laser wakefield acceleration. 
\end{abstract}

% insert suggested PACS numbers in braces on next line
% insert suggested keywords - APS authors don't need to do this
%\keywords{}

%\maketitle must follow title, authors, abstract, \pacs, and \keywords
\maketitle

\section{Introduction}
The safe operation of a particle accelerator requires that the beam be disposed of once it has been used. This is usually achieved using a dense material such as a metal, graphite or water. Such a conventional beam dump can stop even a very high energy electron beam in a relatively short distance. For example, the Large Electron-Positron Collider (LEP) used a $2\mathrm{\ m}$ long aluminium dump to stop a $100\mathrm{\ GeV}$ electron beam \cite{LEP} and the proposed water-based dump for the International Linear Collider (ILC) \cite{ILC_TDR, ILC_dump} is designed to stop a $500\mathrm{\ GeV}$ beam in $11\mathrm{\ m}$. However, the high density of the stopping medium and high power of the beam lead to a number of disadvantages for a conventional dump. Both proton and electron beams lead to the production of radionuclides in the stopping material \cite{Radionuclides_proton,Radionuclides_electron}. The dump must be capable of absorbing the high power of the beam ($18\mathrm{\ MW}$ for the ILC) in a small volume, leading to high power density cooling requirements and high temperatures and pressures \cite{ILC_dump}. The ILC beam dump design would operate at $10\mathrm{\ bar}$ and at a maximum water temperature of $155\mathrm{\ {^\circ} C}$. In the case of a water dump, decomposition generates hydrogen and oxygen gas which must be removed \cite{ILC_TDR,SLAC_dump}. In addition, structural materials may suffer radiation damage and lose strength, a concern for pressure vessels and windows \cite{OECD_radiation_damage,SLAC_dump}. These considerations lead to a conventional beam dump being substantially larger than the length over which they are able to stop their beam may suggest. For instance, the proposed ILC dump will require a pumping station, water tower, catalytic hydrogen-oxygen recombiner, and deionizer sited above ground, connected via pipes to the dump location. A sump is also required to collect any radioactive water that may leak from the dump and ancillary equipment \cite{ILC_dump}.

One proposed alternative to a high density beam dump uses a beam pipe filled with a noble gas at atmospheric pressure, surrounded by iron cladding. With a length of $1000{\mathrm{\ m}}$ the power deposited per unit length is greatly reduced \cite{Leuschner}. Heat can be dissipated by a simple water cooling jacket at atmospheric pressure and radio-activation is reduced compared to the baseline ILC design. However, the disadvantage of this scheme is the extremely long length required to stop a high energy beam and the associated costs of providing space for such a dump.

Another dump scheme, focused on in this paper, uses a plasma wakefield to decelerate a bunch at a high gradient \cite{Wu}. The plasma beam dump minimizes radio-activation by operating at a low density even compared to a gas dump, and potentially allows for the recovery of some of the beam energy as electricity rather than dissipation as heat \cite{Wu}. The high decelerating gradients achievable with high density ultrashort bunches such as those produced by laser wakefield acceleration make plasma beam dumps suitable to complement compact accelerators with compact beam disposal.

In this paper, Section \ref{sec:comparison} compares the plasma beam dump with the conventional beam dump. Section \ref{sec:dump_schemes} discusses the use of modified plasma profiles to improve the performance of a passive plasma beam dump and Section \ref{sec:results} presents particle-in-cell simulation results for a range of plasma beam dump parameters using fixed beam parameters.

\section{Plasma beam dump and conventional beam dump compared}
\label{sec:comparison}
The stopping power, i.e. the average loss of energy $T$ with distance, of an electron in a neutral medium depends on its energy. At high energies, losses are dominated by bremsstrahlung. The critical energy $T_{\mathrm{c}}$ may be defined as the energy at which losses due to bremsstrahlung are equal to losses due to other factors e.g. ionization. The critical energy in MeV is approximated by $T_{\mathrm{c}} = (800{\mathrm{\ MeV}})/(Z+1.2)$ where $Z$ is the atomic number of the stopping material \cite{Proc_shower}. For high-$Z$ materials such as lead or copper, bremsstrahlung dominates at any relevant energy. For lower $Z$ materials such as water, bremsstrahlung is dominant above a few hundred MeV. The stopping power due to radiation is given by\cite{Wu}:
\begin{equation}
  \label{stopping_power}
  {-{{\mathrm{d}T}\over{\mathrm{d}x}}}={Z \alpha} {{4 e^4 n_{\mathrm e}}\over{m c^2}} \left(\gamma - 1\right) \ln \left(183\ Z^{-{1\over 3}}\right)
\end{equation}
where $\alpha$ is the fine structure constant, $m$ is the incident particle mass, $n_{\mathrm{e}}$ is the electron density of the stopping material, $e$ is the elementary charge, $\gamma$ is the relativistic gamma factor and $c$ is the speed of light, with all quantities in CGS units. As long as bremsstrahlung is dominant, the stopping power is linearly proportional to the kinetic energy of the incident particles, $T=\left(\gamma - 1\right) m c^2$.

In a plasma medium, an electron bunch is decelerated by collective oscillations of the electrons in the plasma. The plasma wakefield may be excited by the beam itself in a passive beam dump, or excited by a laser pulse in an active beam dump. The plasma may be preformed or, if the driver is of sufficient intensity, be a neutral gas ionized by the driver itself \cite{Hogan}. A field-ionized plasma would make the passive dump simple and reliable. A passive beam dump has recently been demonstrated experimentally over a short distance using a laser-accelerated bunch \cite{Chou_thesis,Chou}.

The passive dump does however suffer from a major limitation of being unable to decelerate the head of the bunch due to the finite response time of the plasma. This problem can be addressed by the active beam dump, in which the beam is decelerated by the wakefield of a laser pulse \cite{Bonatto}. An active beam dump however relies on the provision of a laser pulse and accurate synchronization. Without either the dump would fail to stop the beam and a backup dump would need to be available.

Recent experimental results have shown that an electron beam can be decelerated by a plasma when initially offset transversely from the plasma column \cite{Adli}. The electron beam is attracted by the charge imbalance created by the beam's transverse fields. In a plasma beam dump employing a pre-formed plasma, this phenomenon would allow the requirements on alignment of the bunch and plasma column width to be relaxed, potentially improving the reliability of the dump.

The highest decelerating gradients for a given plasma density can be achieved in the non-linear regime, where the bunch density exceeds the plasma density. The limit on the maximum decelerating gradient is the wave-breaking field $E_{\mathrm{wb}}$, which depends on the electron plasma frequency $\omega_{\mathrm{p}}$:
\begin{equation}
  \label{wavebreaking_field}
  E_{\mathrm{wb}} = {{m_{\mathrm e} c \,\omega_{\mathrm p}}\over{e}},
\end{equation}
\begin{equation}
  \omega_{\mathrm{p}}=\left({{e^2 n_\mathrm{p}}\over{\epsilon_0 m_e}}\right)^{1\over2},
\end{equation}
where $m_{\mathrm{e}}$ is the electron mass, $\epsilon_0$ is the permittivity of free space and $n_{\mathrm{p}}$ is the plasma electron density, and all quantities are in SI units.

The limit of the wave-breaking field for a plasma density of $10^{24}{\mathrm{\ m^{-3}}}$ can be compared with a copper beam dump for an electron beam of $1{\mathrm{\ GeV}}$. Equation \ref{stopping_power} gives an initial average decelerating gradient of $5.1{\mathrm{\ GeV\ m^{-1}}}$ compared with a wave-breaking field of $96{\mathrm{\ GV\ m^{-1}}}$. The actual decelerating gradient that can be achieved in a plasma depends on the properties of the electron bunch. A short bunch with density higher than the plasma density can achieve a gradient approaching the wave-breaking limit as has been demonstrated experimentally \cite{Hogan,Blumenfeld,Litos,Corde}.

\section{Plasma beam dump schemes}
\label{sec:dump_schemes}
The simplest version of a plasma beam dump is a uniform plasma into which the particle bunch to be decelerated propagates. The head of the bunch will experience no decelerating field, while some part of the bunch will experience a maximum decelerating field. After some time the part of the bunch that experiences the maximum field will become non-relativistic and will fall behind the rest of the bunch until it reaches an accelerating region of the wakefield. The portion of the bunch will then absorb energy from the wakefield and be re-accelerated. This leads to the rate of energy loss of the bunch dramatically decreasing after a saturation length $L_{\mathrm{sat}}$, as a substantial proportion of the energy lost is reabsorbed. The saturation length for a beam of initial energy $T_0$ is approximately the propagation length at which the maximum decelerating gradient $E_{\mathrm{dec}}$ decelerates a portion of the beam to non-relativistic velocity:
\begin{equation}
  L_{\mathrm{sat}} \approx {{T_0}\over{e E_{\mathrm{dec}}}}.
\end{equation}
Wu {\it et al.} proposed to use a structured plasma consisting of a series of foils starting after $L_{\mathrm{sat}}$ to absorb the low energy particles and prevent them from being re-accelerated \cite{Wu}. The presence of thin foils in the path of a high power beam, and the potential for high temperatures and electric fields in the plasma may lead to damage to the foils. A scheme that achieves the same result using a plasma-only decelerating medium may be attractive.

\subsection{Varying plasma density}
\begin{figure}[htb]
  \centering
  \includegraphics[width=70mm]{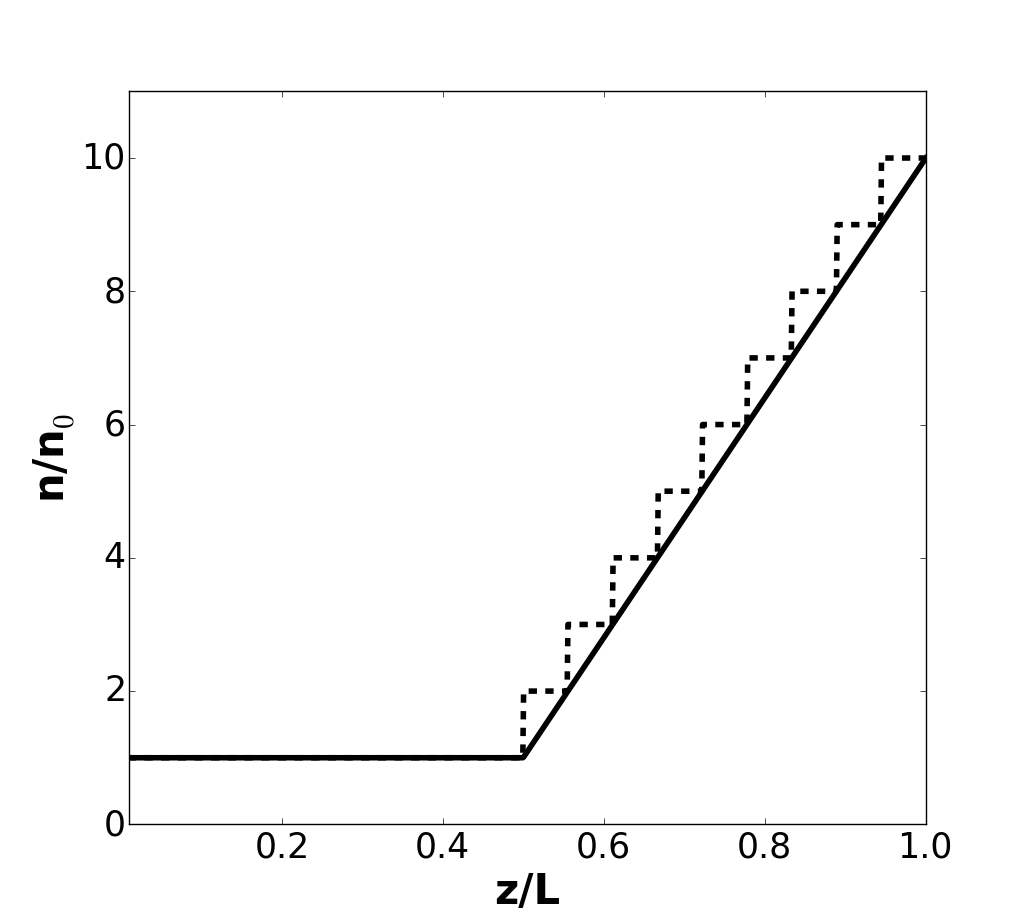}
  \caption{Plot illustrating stepped (dashed line) and linear gradient (solid line) plasma density schemes. The density is constant over the saturation length $L_{\mathrm{sat}}$ and then increases to ten times the initial density over the remaining length.}
  \label{density_diagrams}
\end{figure}

As an alternative to the use of foils to absorb low energy particles, the decelerated particles can be removed from the accelerating region of the wakefield by defocusing. This can be achieved by increasing the plasma density once the bunch has travelled over the saturation length. The plasma wavelength $\lambda_{\mathrm{p}}$ is related to the plasma electron density $n_{\mathrm{p}}$ by:
\begin{equation}
  \label{plasma_wavelength}
  \lambda_{\mathrm p} = {2 \pi c} \left({\epsilon_0 m_{\mathrm e}}\over{e^2 n_{\mathrm p}}\right)^{1 \over 2} \approx 3.3\times 10^7/ \sqrt{n_{\mathrm{p}}}
\end{equation}
where all quantities are in SI units. As the density increases the plasma wavelength decreases, effectively shifting the bunch within the wakefield. Half of one plasma wavelength behind the drive bunch is the region of highest on-axis electron density. If the plasma density is increased, decelerated particles will pass through a strong defocusing region and be removed from the axis. This will prevent their re-acceleration. For a stepped plasma, the change in plasma wavelength is instantaneous and the decelerated particles do not need to pass through the accelerating region. For a gradual plasma density increase the decelerated particles will gain energy in the accelerating region prior to being defocused. Figure \ref{density_diagrams} shows a diagram of the stepped and gradient plasma density schemes. The rate of change with position of the plasma wavelength can be calculated for a given plasma profile, by taking the derivative of $\lambda_{\mathrm p}$ (Equation \ref{plasma_wavelength}) with respect to $z$. For a linearly increasing plasma density from initial density $n_{\mathrm{i}}$ to a final density $n_{\mathrm{f}}$ over a length $L$:
\begin{equation}
  \label{plasma_wavelength_change}
  {{\mathrm{d}\lambda} \over {\mathrm{d}z}}={{\pi c\,e^2 n_{\mathrm{f}}}\over{\epsilon_0 L m_{\mathrm{e}}}}\left[{e^2 n_{\mathrm{i}}} \left(1+{{{{n_{\mathrm{f}}}}\over{n_{\mathrm{i}}}}{z \over L}}\right)\over{\epsilon_0 m_{\mathrm{e}}}\right]^{-{3 \over 2}}.
\end{equation}

\begin{figure}[htb]
  \centering
  \includegraphics[width=70mm]{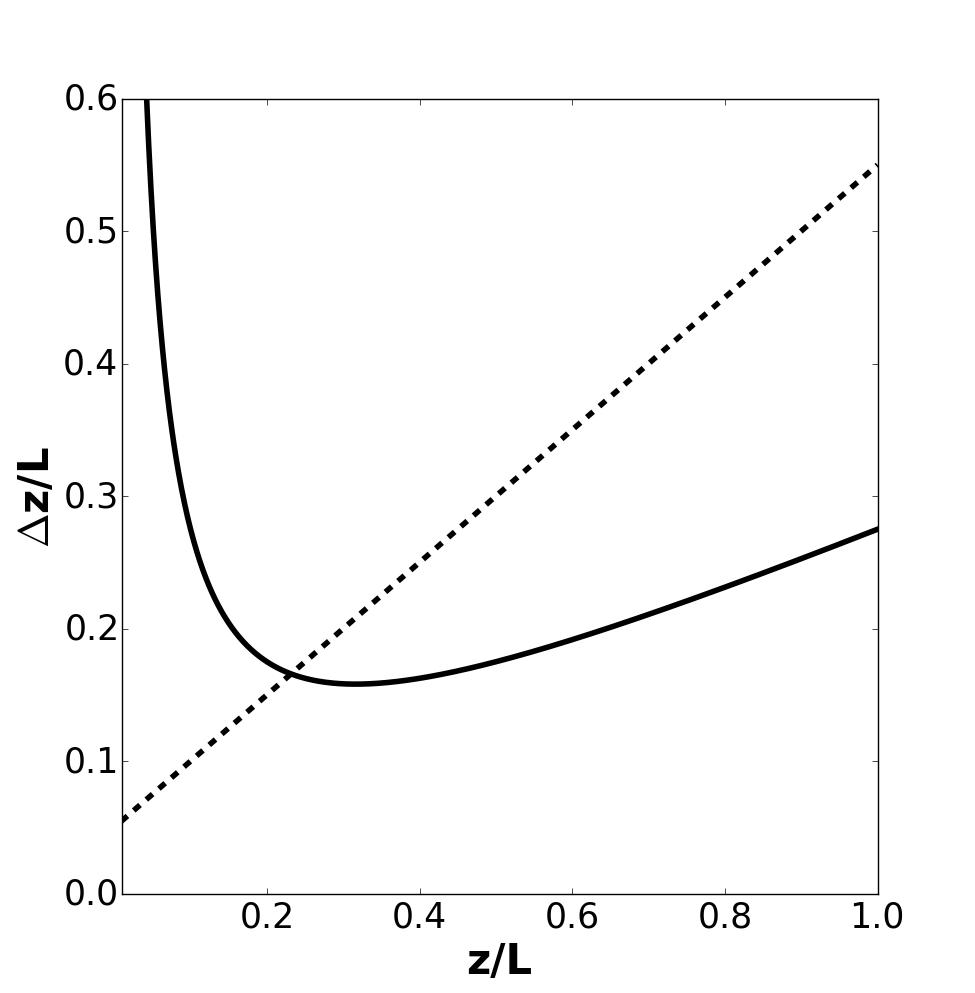}
  \caption{Length over which a low-energy particle is accelerated for linear (dashed line) and quadratic (solid line) plasma density increases over a distance $L$. The singularity in the quadratic case is a result of the assumption that the rate of plasma wavelength change is constant for each data point, and this is zero at $z=0$.}
  \label{gradient_energy_gain}
\end{figure}

In the linear regime, the defocusing region is located $\lambda_{\mathrm{p}} /4$ behind the maximum decelerating region. The propagation distance $\Delta z$ over which the plasma wavelength changes by $1/4$ can be estimated assuming that the rate of change of plasma wavelength is constant, and the energy gain $\Delta T$ is the average accelerating field $E_{\mathrm{acc}}$ multiplied by the propagation distance.
\begin{equation}
  \label{delta_x}
  \Delta T = E_{\mathrm{acc}} \Delta z = E_{\mathrm{acc}} {{\lambda_0}\over 4}\left({{{\mathrm{d}\lambda}} \over {\mathrm{d}z}}\right)^{-1},
\end{equation}
where $\lambda_0$ is the initial plasma wavelength at a given position $z$. The  more rapid the change in plasma density, the less energy will be gained by the decelerated particles, however the density has to remain low enough to be achievable and to generate a high decelerating field. In this study the density was increased by a factor of ten over the plasma length. Linear and quadratic plasma density changes were considered. Figure \ref{gradient_energy_gain} shows a plot of the distance required for the plasma wavelength to change by ${{\lambda_0}/4}$. A quadratic density profile maintains the rate of plasma wavelength change at a higher level compared to a linear density ramp. This will lead to reduced re-acceleration and thus a more effective beam dump.

\section{Simulation results}
\label{sec:results}
Two-dimensional simulations of passive beam dump schemes were carried out using the explicit particle-in-cell (PIC) code VSim \cite{VSim}. Bunch parameters were chosen to represent a bunch that can be generated routinely by laser wakefield acceleration \cite{Karsch,Osterhoff,Buck,Malka}. The bunch has an rms length of $7.5{\mathrm{\ \upmu m}}$, rms radius of $20{\mathrm{\ \upmu m}}$ and charge of $100{\mathrm{\ pC}}$. The total energy of the bunch is $0.025\mathrm{\ J}$ corresponding a bunch which may be generated by a modest laser pulse of $0.25\mathrm{\ J}$ assuming $10\%$ laser to bunch efficiency \cite{Malka_2006}.  A moderate energy of $250{\mathrm{\ MeV}}$ allows the simulation length to be kept short. A $25{\mathrm{\ cm}}$ plasma length allows the deceleration to saturate and for modified density schemes to be studied. An initial plasma density $n_{\mathrm{i}}$ of $2\times 10^{23} {\mathrm{\ m^{-3}}}$ was used, for a range of step lengths and for linear and quadratically increasing plasma density after the saturation length. For the gradient plasma schemes the density was increased by a factor of 10 starting from the saturation length and ending at the end of the plasma. The step schemes increased the plasma density by $n_{\mathrm{i}}$ for each step. The step length refers to the length of the flat plasma density between each step.

\begin{figure}[htb!]
  \centering
  \includegraphics[width=70mm]{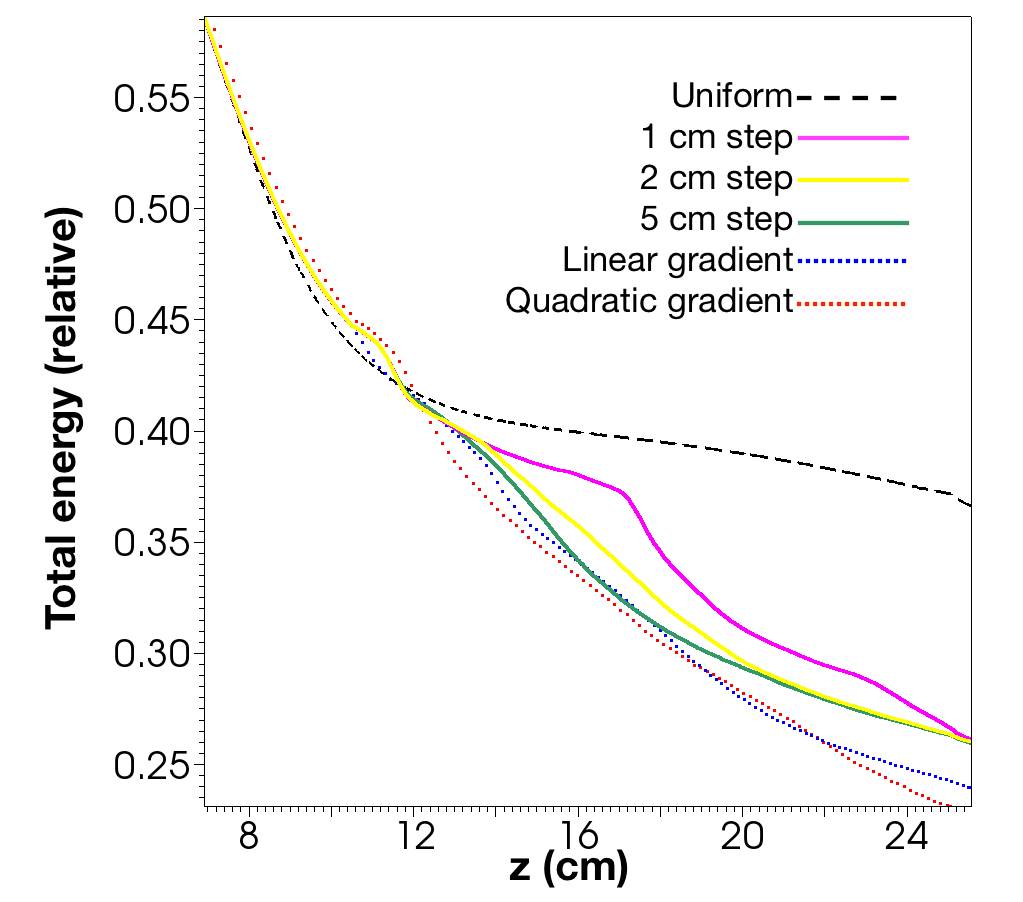}
  \caption{Energy loss with distance for uniform, stepped and gradient plasma density profiles. Prior to approximately $z=10\mathrm{\ cm}$ all profiles show the same constant decelerating gradient.}
  \label{energy_plots}
\end{figure}

\begin{figure}[htb]
  \centering
  \subfloat[]{
    \includegraphics[width=75mm]{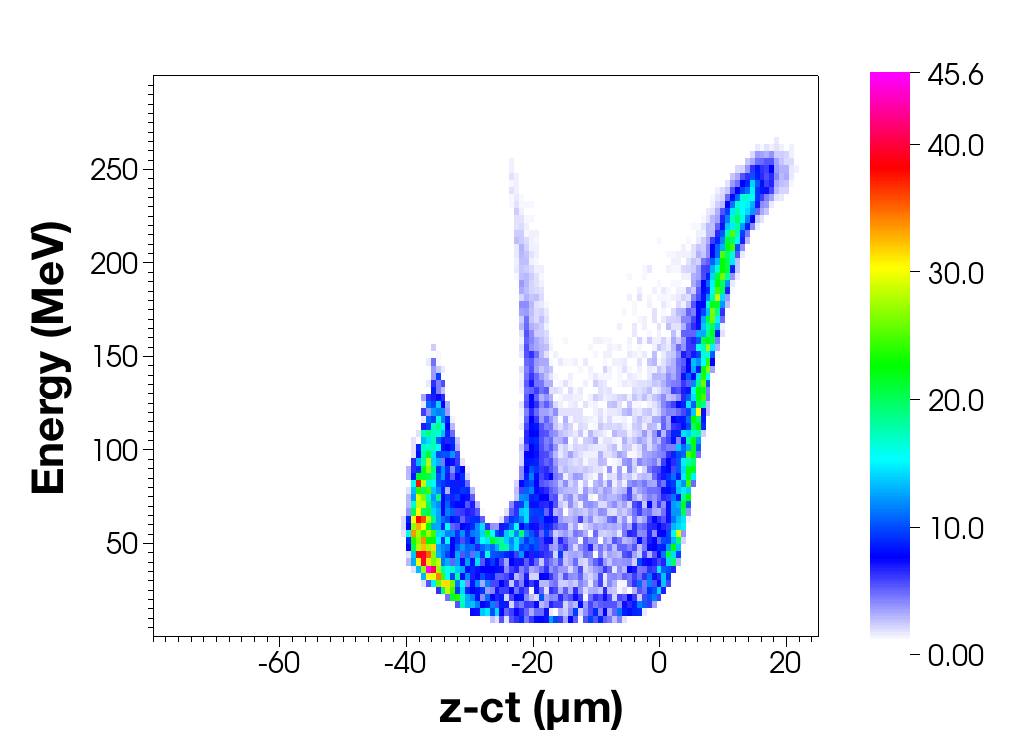}
    \label{uniformLongPhaseSpace}
  }
  \subfloat[]{
    \includegraphics[width=75mm]{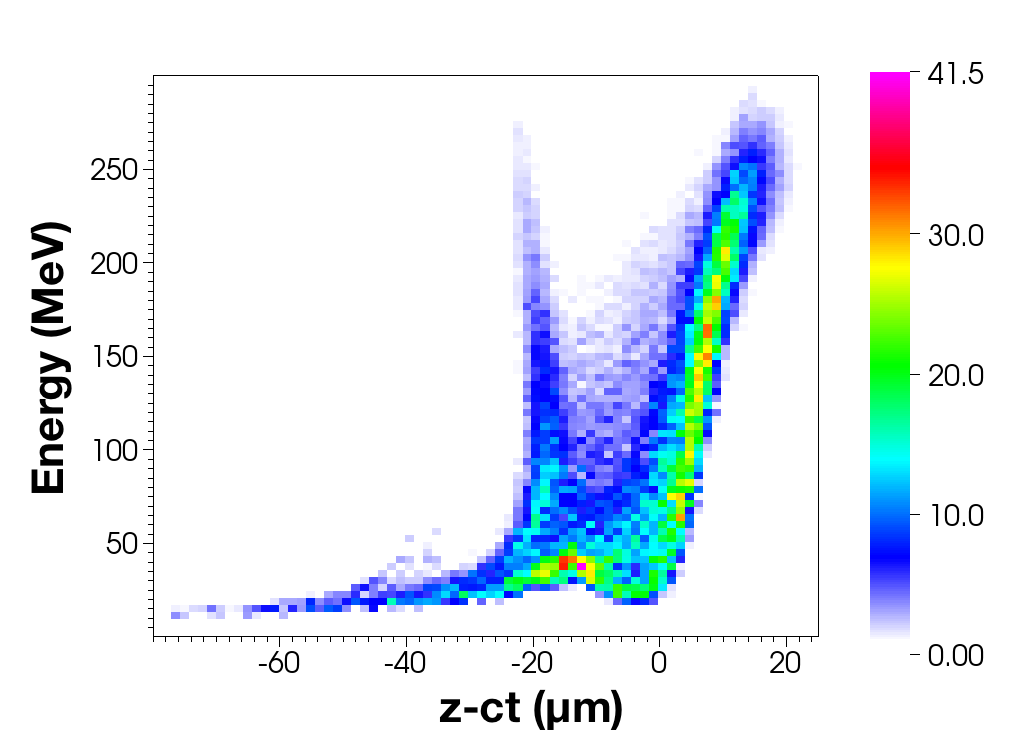}
    \label{stepLongPhaseSpace}
  }
  \caption{Longitudinal phase space histogram at $z=16.3{\mathrm{\ cm}}$ for a uniform plasma (a) and a linear gradient plasma profile (b). The energy scale corresponds to $\gamma/{m_{\mathrm{e}}c^2}$ and as such is not accurate for non-relativistic velocities. The color scale gives the sum of macroparticle weight for each bin.}
  \label{longPhaseSpaceComparison}
\end{figure}

\begin{figure}[htb]
  \centering
  \subfloat[]{
    \includegraphics[width=75mm]{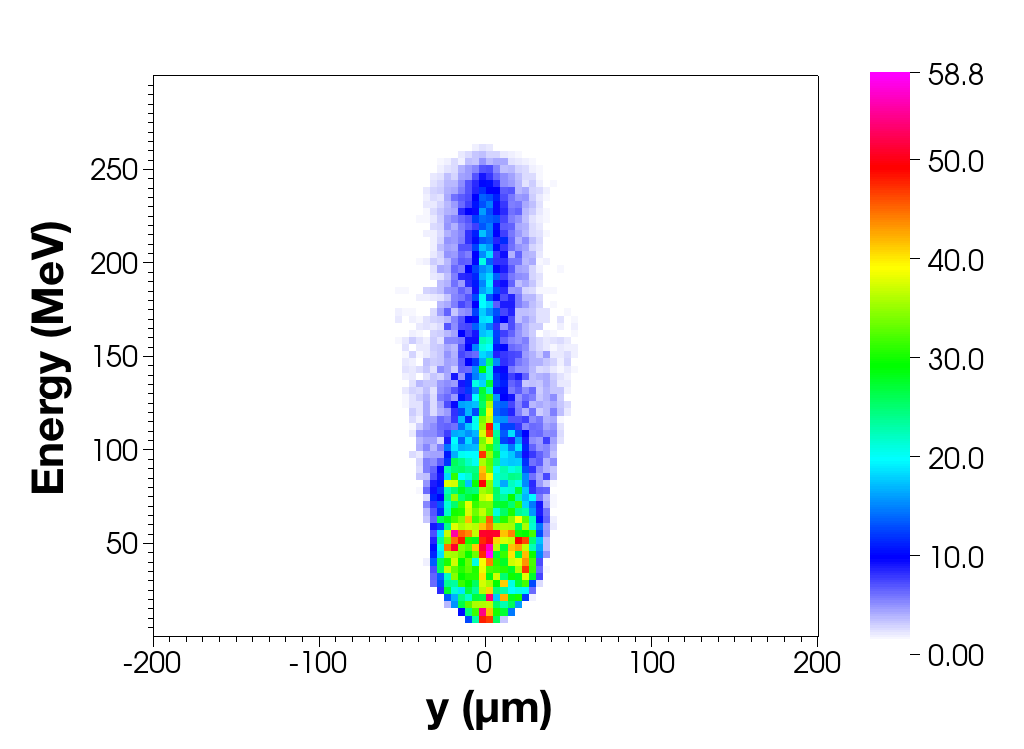}
    \label{uniformTransverse}
  }
  \subfloat[]{
    \includegraphics[width=75mm]{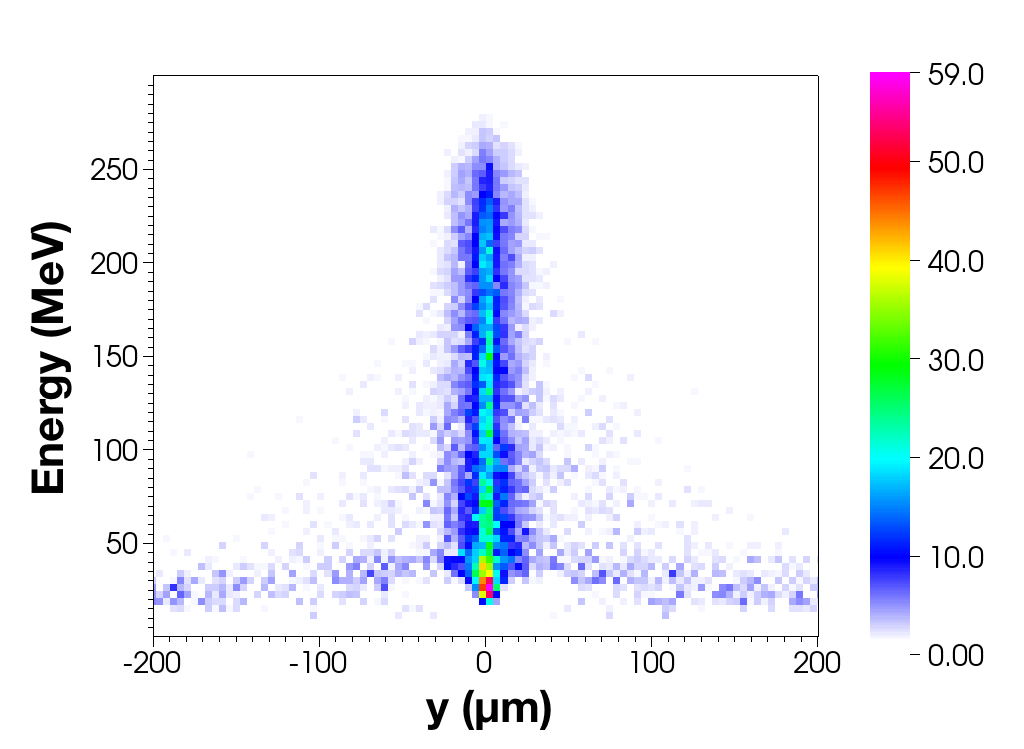}
    \label{stepTransverse}
  }
  \caption{Histogram of energy vs. transverse coordinate at $z=16.3{\mathrm{\ cm}}$ for a uniform plasma (a) and a linear gradient plasma profile (b). The color scale gives the sum of macroparticle weight for each bin.}
  \label{transPhaseSpaceComparison}
\end{figure}

\begin{figure}[htb]
  \centering
  \includegraphics[width=70mm]{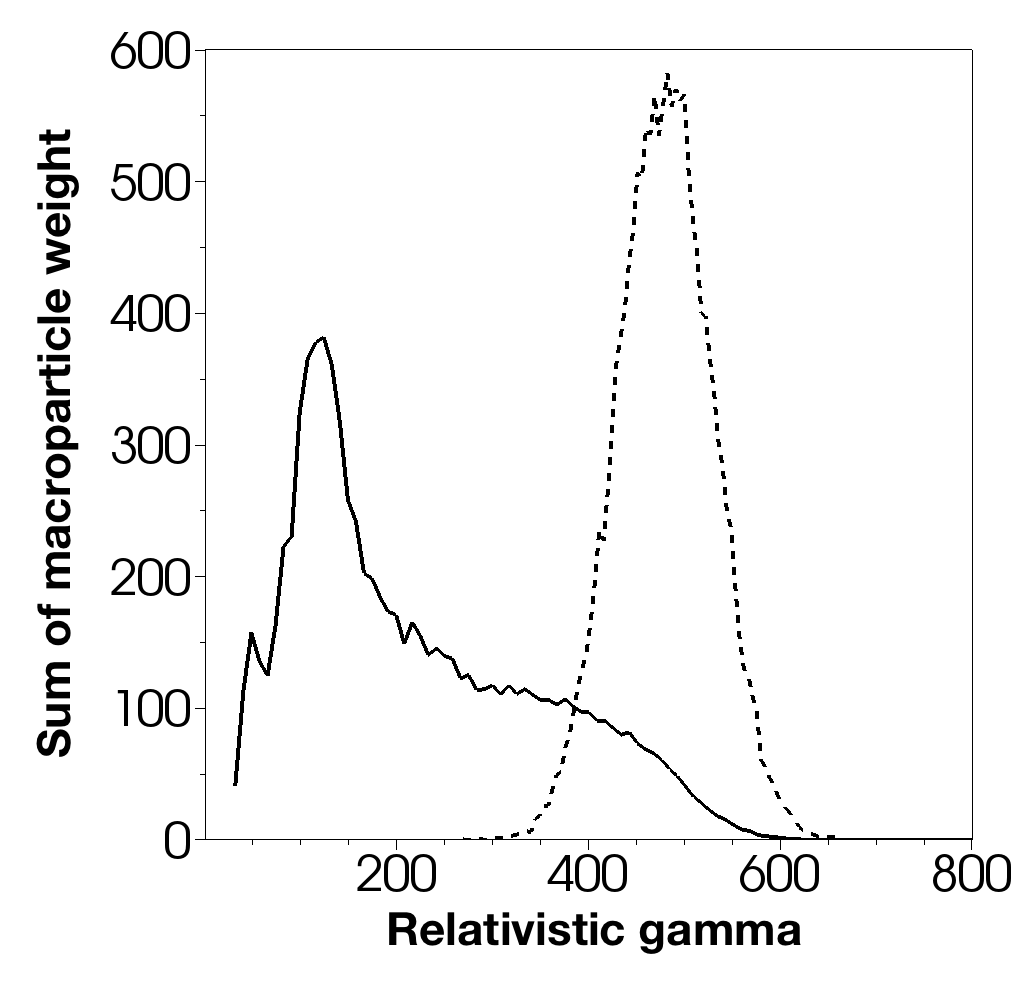}
  \label{spectrum_gradient}
  \caption{Histogram of $\gamma$ of the electron bunch at $z=0$ (dashed line) and after $25{\mathrm{\ cm}}$ (solid line) for a linear gradient plasma profile. The $y$-scale is the sum of macroparticle weight in each bin. 100 equally-sized bins were used.}
  \label{gradient_energy_spectrum}
\end{figure}

Figure \ref{energy_plots} shows the change in total beam energy for different dump schemes. The gradient scheme proved to provide the greatest energy loss over $25 {\mathrm{\ cm}}$. Figure \ref{longPhaseSpaceComparison} shows the longitudinal phase space for a uniform plasma and a $1\mathrm{\ cm}$ stepped plasma profile. In each plot the bunch has propagated $16.3{\mathrm{\ cm}}$ in the plasma, some distance beyond the saturation length of approximately $10\mathrm{\ cm}$. It can be seen that significantly less charge is re-accelerated in the case of the gradient profile. In the region $z-ct<-30{\mathrm{\ \upmu m}}$, outside the extent of the initial bunch, at $z=16.3{\mathrm{\ cm}}$ there was found to be $25{\mathrm{\ pC}}$ of charge re-accelerated to greater than $30{\mathrm{\ MeV}}$. This compares with only $0.6{\mathrm{\ pC}}$ in the gradient plasma density case. Figure \ref{transPhaseSpaceComparison} shows a plot of energy and transverse position at the same propagation distance. The lowest energy part of the bunch has been defocused while the higher energy particles have not been affected. The defocused particles are limited to less than approximately $50\mathrm{\ MeV}$. A comparison of the initial and final energy spectra for the linear ramp case is shown in Figure \ref{gradient_energy_spectrum}. Although the peak energy of the bunch remains largely unchanged, the intensity of particles at the initial central energy has been reduced by a factor of 10. The relativistic $\gamma$ at the peak intensity of the final bunch corresponds to an energy of approximately $75\mathrm{\ MeV}$.

\section{Conclusion and outlook}
\label{sec:conclusion}
Simulation results show that a short, moderate charge electron bunch can lose a large fraction of its energy in a $25\mathrm{\ cm}$ plasma. Stepped and gradient plasma profiles are capable of improving energy loss and provide an alternative to the previously proposed foil scheme. Gradient plasma profiles were found to be most effective in improving energy loss, however there was relatively little difference between the linear and quadratic plasma profiles. The advantage of the gradient scheme suggests that the energy gain of non-relativistic particles while the plasma wavelength changes is not significant for the parameters used. However, this may not be the case when bunch parameters are such that the accelerating gradient is very large. The achievability of the modified plasma density profiles will depend on the technology used for the source, which will in turn depend on the beam and plasma parameters. Such considerations will be important if such a passive plasma beam dump is to be experimentally tested in the future.

Plasma beam dumps show great promise in both providing compact deceleration to complement high-gradient novel accelerators and in reducing the complexity of beam dumps in conventional accelerators. Although passive plasma beam dumps are not capable of decelerating the head of the bunch, the rapid reduction in total beam energy would allow for a conventional beam dump to absorb the remaining energy with greatly reduced radio-activation and cooling requirements.

% If you have acknowledgments, this puts in the proper section head.
\begin{acknowledgments}
  This work was supported by the Cockcroft Institute Core Grant and the STFC.
  The authors gratefully acknowledge the computing time granted on the supercomputer JURECA at J\"ulich Supercomputing Centre (JSC).
\end{acknowledgments}

% Create the reference section using BibTeX:
\bibliography{beamDump_aip_v3_bibliography}

\end{document}